\documentclass[aps,twocolumn,showpacs,preprintnumbers,amsmath,amssymb,floatfix]{revtex4-2}
\usepackage{graphicx}
\usepackage{epsfig}
\begin{document}
\def\be{\begin{equation}}
\def\ee{\end{equation}}
\def\bearr{\begin{eqnarray}}
\def\eearr{\end{eqnarray}}
\def\tc{$T_c~$~}
\def\tcl{$T_c^{1*}~$~}
\def\c2{ CuO$_2~$~}
\def\ruo{ RuO$_2~$~}
\def\lsco{LSCO~}
\def\bi{bI-2201~}
\def\tl{Tl-2201~}
\def\hg{Hg-1201~}
\def\sro{$\rm{Sr_2 Ru O_4}$~}
\def\rc{$RuSr_2Gd Cu_2 O_8$~}
\def\mgb{$MgB_2$~}
\def\pz{$p_z$~}
\def\ppi{$p\pi$~}
\def\sqo{$S(q,\omega)$~}
\def\tperp{$t_{\perp}$~}
\def\he4{${\rm {}^4He}$~}
\def\ags{${\rm Ag_5 Pb_2O_6}$~}
\def\nxcob{$\rm{Na_x CoO_2.yH_2O}$~}
\def\lsco{$\rm{La_{2-x}Sr_xCuO_4}$~}
\def\lco{$\rm{La_2CuO_4}$~}
\def\lbco{$\rm{La_{2-x}Ba_x CuO_4}$~}
\def\half{$\frac{1}{2}$~}
\def\thalf{$\frac{3}{2}$~}
\def\tst{${\rm T^*$~}}
\def\tch{${\rm T_{ch}$~}}
\def\jeff{${\rm J_{eff}$~}}
\def\nbc{${\rm LuNi_2B_2C}$~}
\def\cabc{${\rm CaB_2C_2}$~}
\def\nboo{${\rm NbO_2}$~}
\def\voo{${\rm VO_2}$~}
\def\nip{$\rm LaONiP$~}
\def\nisb{$\rm LaONiSb$~}
\def\nibi{$\rm LaONiBi$~}
\def\fep{$\rm LaOFeP$~}
\def\cop{$\rm LaOCoP$~}
\def\mnp{$\rm LaOMnP$~}
\def\fesb{$\rm LaOFeSb$~}
\def\febi{$\rm LaOFeBi$~}
\def\efeas{$\rm LaO_{1-x}F_xFeAs$~}
\def\hfeas{$\rm La_{1-x}Sr_xOFeAs$~}
\def\hSfeas{$\rm Sm_{1-x}Sr_xOFeAs$~}
\def\hCefeas{$\rm Ce_{1-x}Sr_xOFeAs$~}
\def\feas{$\rm LaOFeAs$~}
\def\Ndfeas{$\rm NdOFeAs$~}
\def\Smfeas{$\rm SmOFeAs$~}
\def\Prfeas{$\rm PrOFeAs$~}
\def\refeas{$\rm REOFeAs$~}
\def\refesb{$\rm REOFeSb$~}
\def\refebi{$\rm REOFeBi$~}
\def\ttog{$\rm t_{2g}$~}
\def\fese{$\rm FeSe$~}
\def\fete{$\rm FeTe$~}
\def\eg{$\rm e_{g}$~}
\def\dxy{$\rm d_{xy}$~}
\def\dzx{$\rm d_{zx}$~}
\def\dzy{$\rm d_{zy}$~}
\def\dxsq{$\rm d_{x^{2}-y^{2}}$~}
\def\dzsq{$\rm d_{z^{2}}$~}
\def\LAO{$\rm LaAlO_3$~}
\def\STO{$\rm SrTiO_3$~}
\def\htwos{$\rm H_2 S$~}
\def\hthrees{$\rm H_3 S$~}
\def\htwo{$\rm H_2$~}
\def\silane{$\rm Si H_4$~}
\def\lahten{$\rm LaH_{10}$~}
\def\lahtwo{$\rm LaH_{2}$~}
\def\lahthree{$\rm LaH_{3}$~}
\def\methane{$\rm CH_{4}$~}
\def\yhtwo{$\rm YH_{2}$~}
\def\yhthree{$\rm YH_{3}$~}
\def\yhten{$\rm YH_{10}$~}
\def\luhthree{$\rm LuH_{3}$~}
\def\luhthreenx{$\rm LuH_{3-x}N_x$~}
\def\lunxhx{$\rm LuN_{1-x}H$~}
\def\hminus{$\rm H^{-}$~}
\def\hplus{$\rm H^+$~}
\def\yhthreex{$\rm YH_{3-x}$~}
\def\lahthreex{$\rm LaH_{3-x}$~}
\def\cahsix{$\rm Ca^{}H_6$~}
\def\abohr{$\rm a_B$~}
\def\abohrstar{$\rm a_B^*$~}
\def\onestars{$\rm 1^*s^1$~}
\def\height{$\rm H_8$~}
\def\silane{$\rm SiH_{4}$~}
\def\lunh{$\rm LuNH$~}
\def\PbxCuxApatite{$\rm Pb_{9+x}Cu_{1-x}(PO_4)_6O$~}
\def\pbap{$\rm Pb_{10}(PO_4)_6O$~}
\def\agpbo{$\rm Ag_5Pb_2O_6$~}
\def\pb2plus{$\rm Pb^{2+}$~}
\def\O2minus{$\rm O^{2-}$~}
\def\phospate3plus{$\rm (PO_4)^{3-}$~}
\def\cu2plus{$\rm Cu^{2+}$~}
\def\cuzero{$\rm Cu^0$~}

\title{Broad Band Mott Localization is all you need for Hot Superconductivity:\\
Atom Mott Insulator Theory for Cu-Pb Apatite}
 
\author{ G. Baskaran}
\affiliation
{The Institute of Mathematical Sciences, C.I.T. Campus, Chennai 600 113\\
Department of Physics and QuCenDiEM, Indian Institute of Technology, Chennai 600 036, India \&\\
Perimeter Institute for Theoretical Physics, Waterloo, ON, Canada}

\begin{abstract}
A hypothetical non-dimerized Cu chain in equilibrium is a spin-\half atom Mott insulator (AMI), eventhough its band width is high $\sim$ 10 eV. This RVB reservoir has a large exchange coupling J $\sim$ 2 eV. This idea of, \textit{broad band Mott localization} was used by us in our earlier works, including prediction of high Tc superconductivity in doped graphene, silicene and a theory for hot superconductivity reported in Ag-Au nanostructures (TP 2008). In the present work we identify possible random AMI subsystems in Cu-\pbap and develop a model for reported hot superconductivity (LKK 2023). In apatite structure, network of interstitial columnar spaces run parallel to c-axis and ab-plane. They accomodate excess copper, as \cuzero clusters, chains and planar segments. They are our emergent AMI's. Electron transfer from AMI's to insulating host, generates strong local superconducting correlation, via phyics of doped Mott insulator. Josephson coupling between doped AMI's, establishes hot superconductivity. A major Challenge to superconducting order in real material is competing insulating phases - valence bond solid (spin-Peirels)-lattice distortions etc. AMI theory points to ways of making the \textit{elusive superconductivity} palpable. We recommend exploration of hot superconductivity in the rich world of minerals and insulators, via metal atom inclusion.
\end{abstract} 

\maketitle

\section*{Introduction}

A milestone discovery of high Tc superconductivity in cuprates, by Bednorz and Muller \cite{BednorzMuller1986}  in 1986 heralded serious attempts to discover new superconductors, possibly with higher Tc's. Resonating valence bond theory of high Tc superconductivity, a response to understand cuprates, pioneered by Anderson \cite{PWA1987,BZAandOthers,KRS,KotliarDwave,PlainVanila,GBIran} created a fertile ground for new physics, variety of quantum spin liquids \cite{WenBook,TaiKaiSpinLiquid} etc. Doped spin-\half Mott insulators became a playground for high Tc superconductivity and offered new hope, inspite of competing orders \cite{GB5FoldWay}.

Announcement of hot superconductivity in certain hydrides \cite{DiasLuNH,LaHx400}, Ag-Au nanostrucutes \cite{ThapaPandey} and more recently, Cu-\pbap \cite{LK99} has excited the community. Historically there have been claims of hot superconductivity, named \textit{unstable and elusive superconductors}\cite{Kopelevich}. 

In the past, inspired by possibility of broad band Mott localization we have developed theories of superconductivity in doped graphene-graphite \cite{GBMgB2,AnnaDoniach,PathakShenoyGB}, silicene \cite{GBSilicene}, B doped diamond \cite{GBDopedDiamond}, K$_3$ p-Terphenyl \cite{GBpTerphenyl}, Hydride \cite{GBHydride} and Ag-Au nanostructures \cite{GBAgAu1andIJMP,GBAgAu2019}. Here atom Mott insulators (AMI)and near Mott localization in low dimensional monoatom assemblies play special role.

In what follows, we propose a model for possibile hot superconductivity in Cu-\pbap system, based on our AMI theory \cite{GBAgAu2019,GBAgAu1andIJMP} applied to reported superconductivity in Ag-Au nanostructures.   Our arguments are rather qualitative. However,  they enumerate rich possibilities, likely to have implications for future studies.

In \pbap, interstitial sites of Pb octahedral columns are half empty. In addition, a network of c-axis interstitial columns and ab-plane interstital columns are present in the ionic-covalent insulating matrix. Our hypothesis is that excess Cu gets accommodated as a random network of weakly connected clusters, chains and 2D patches of \cuzero in the network of interstitial columns. These are the weakly coupled random network of emergent AMI's

Difference in electronegativity induces electron transfer from Mott insulating \cuzero AMI subsystem to the host insulator. It is a good assumption that transferred electrons get localized in the random envoronment of the disturbed host. A random network of hole doped Mott insulator emerges. In doped Mott insulator, resonating singlets generate strong Cooper pair correlation. In the pair tunneling mechanism of superconductivity \citep{WHA}, because of spin-charge decoupling present in the Hubbard chains, single electron tunneling between chains is blocked at low energy scales. However, pair tunneling is not blocked. This Josephson type pair tunneling selfconsistently builds a 3D superconducting state\cite{WHA,ZKEmery}. 

This weakly coupled doped reference Mott insulator, a coupled t-J model of chains and clusters explains observed hot superconductivity, for our estimates of model parameter for Cu-Pb apatite. A major Challenge to our mechanism of superconductivity is competing (insulating) orders such as valence bond solid (spin-Peirels) and structural distortions, within AMI subsystem.

Interestingly \textit{anomalous diamagnetism and ferromagnetism at room temperatures} have been reported in monoatomic Cu, Ag and Au nonosystems
 \cite{CuDiaMSRao,ImryDiamagnetism,AuDiaNanorod,AuDiaNanorodRhee,AgCuCompositeDia,AgRingDia,AgAuFM} 
This underscores importance of electron electron Coulomb interaction effects in broad band monovalent atom systems in low dimensions.

In the present paper we focuss on possible doped Mott insulator physics that has proven validity in the context of single band system such as cuprates and other strongly correlated superconductors, consistent with our earlier proposal of `Five fold way to new superconductors' \cite{GB5FoldWay} and emergent Mott insulators in effectively atomic Mott systems \cite{GBMgB2,PathakShenoyGB,GBSilicene,GBDopedDiamond,GBpTerphenyl}.

Our article is organized as follows. First we present arguments for how isolated bare narrow bands (for example in the \cu2plus substitution scenerio) will not give us scale of Tc reported and a need for broad band Mott localization. After a summary of phenomenology of superconducting Cu-Pb Apatite, we present a structural-electronic model for Cu subsystem in Cu-Pb Apatite. Emergence of weakly coupled effectively Mott insulating cluster, chains and patches of Cu$^0$ AMI's is presented (4s orbital are weakly hybridized with orbitals of surrounding ions). Then we present inter\textit{ cluster, chain, 2D patches} pair tunneling mechanism, coupled t-J chain model relevant for weakly coupled doped Mott insulating chains and estimates of superconducting Tc's. Obstacles to reach hot superconductivity, via competing orders prevalent in low dimensional Mott insulating systems and electron lattice coupling is discussed next.

\section*{Why Broad Band Mott Localization ?}
Many mechanisms for high Tc superconductivity exists. One of them is the doped spin-\half Mott insulator, RVB mechanism. In this route an effective single band close to half filling plays a key role. Known doped Mott insulators, for example high Tc cuprates have a bare band width of about 2 eV and singlet exchange coupling is about 0.15 eV. Maximum Tc possible in cuprate system (from theory point of view) is about 200 K - it is never realized in practice, because of unavoidable competing orders. In organic superconductors (which could be viewed as self doped Mott insulators) bands are about 0.5 eV wide; experimentlly seen maximum Tc is about 12 K. 

In cuprates the physics is not one of emergence of heavy quasiparticles at the Fermi leve, but that of celebrated spin-charge decoupling, failure of Fermi liquid state and absence of Fermi liquid quasi particles at the Fermi level. It is interesting to recall emergence of heavy effective mass of Fermi liquid quaiparticles and an emergent flat band at the Fermi level in heavy Fermions. Here bare band width of conduction electrons is very wide. However, Kondo exchange with localized spins generates, at low energy scale an effective narrow quasi particle band (built out of Kondo resonances) at the Fermi level. This strongly correlated state has superconductivity. However, scale of Tc is limited by Kondo temperature in an exponential fashion. 

If one assumes, as the authors of LK-99 have suggested, replacement of one chain of Pb$^{2+}$ by \cu2plus, one should get a narrow and isolated bare half filled band at the Fermi level. This is because of large Cu-Cu separation, and an unfriendly envoronment (as explained below)for \cu2plus. Conequently, as recent ab-initio calculations \cite{LK99Abinitio} show, one gets isolated bare narrow bands. Correlation effects will further narrow the band. When we estimate Tc, using doped Mott insulator physics, we get a Tc not exceeding 10K. Even this Tc will be reduced further by disorder effects and singlet localization via lattice distortion.

So we are forced to think about an alternative. Here, broad band Mott localization comes in handy and could push Tc's to the scales observed. 

\section*{Superconducting Phenomenology of LK99}
\pbap, an ionic insulator is a member of a large family of apatite minerals, with a nominal charge state Pb$^{2+}_{10}$(PO$_4)^{3-}_6$O$^{2-}$. The authors \cite{LK99} state that replacement of about a tenth of Pb$^{2+}$ ions by Cu$^{2+}$ ions results in an insulator to metal transition. They report Meissner signal, vanishing resistivity, levitation. and Tc exceeding 400 K. It is important to point out that their structural analysis using XRays, shows a minor phase of Cu$_2$S. Defect concentrations, nature of structural irregularies, size of crystallites are not reported. 

They also sketch a mechanism of superconductivity based on substitution of \pb2plus by \cu2plus in the c-axis  chains of Pb. Very recent ab-initio calculation\cite{LK99Abinitio} shows presence of well isolated narrow 3d band of \cu2plus. In our estimate the entire isolated band is too narrow and could support only very low Tc superconductivity, via Mott localization physics, as briefly explained in the next section.

Recent works \cite{LK99Awana,LK99Chinese} report succesfull synthesis of LK99; their search for superconductivity continues.

\section*{A structural model and reference electronic state for LK99} 
What are the charge state of added Cu in LK-99 ? In general, local solid state and quantum chemistry and energetics dictate stable and possible metastable charge states of added atom. Cu is known to become Cu$^{2+}$ ion, when there is covalent mixing with oxygen atoms, as in copper complexes, solid CuSO$_4$, CuO and cuprates. 

Cu$^{2+}$ in principle can replace Pb$^{2+}$ in the Pb chain. However, the trigonal prismatic local envoronment provided by corners of 6 surround phosphate tetrahedra, is not conducive for Cu$^{2+}$ ion replacement. Ionization energy of 4s electron of Cu is high, about 7.7 eV.  It is also likely that lone pair electrons of Pb$^{2+}$ has an important role in not selecting Cu$^{2+}$ state. When one replaces \pb2plus ions by \cu2plus, according to very recent ab-initio studies \cite{LK99Abinitio} one gets narrow and well isolated Cu 3d band at the Fermi velvel (even without correlation effects). As mentioned earlier superconducting Tc's given by electronic mechanism in this bare and isolated narrow band is very small. 

In what follows we present possibility of excess Cu atoms getting accommodated as clusters, chains and 2D patches of \cuzero and becoming a source of atom Mott insulator (AMI) and broad band Mott localization. To appreciate this, we discuss the structure first. Apatite minerals have complex structures and large unit cells. Nature has used millions of years to arrive at a pretty periodic structure in \pbap. 

Pairs of \phospate3plus tetrahedra stack to form columns, which run parallel to c-axis and form a Kagome lattice. \pb2plus ions create c-axis chains and form a hexagonal lattice.  Pb$^{2+}_6$ octahedra share faces and create c-axis columns of triangular lattice. Half of the interstitial sies of the octahedra are occupied by \O2minus ions. (Triangular and Kagome lattices are formed respectively by centers of elementary hexagon and centers of nearest neighbor bond of honeycomb lattice). Asymmetric pairing of Phosphate tetrahedra adds possibility of chirality orders in apatite. 

The half empty interstitials in Pb octahedral columns in real systems are likely to have disordered O$^{2-}$ occupantion, leaving pieces of empty interstitial columns. In addition network of c-axis interstitial columns and ab-plane interstital columns are present in the ionic-covalent netowrk. Further minerals, including apatites with tetrahedral moieties are known to be good ionic (for example O$^{2-}$) conductors \cite{ApatiteIonicCond}. Since ionic radii of O$^{2-}$ ion \cuzero are very close, we expect our proposed channel network to accomodate \cuzero atoms.

Our hypothesis is that excess Cu gets accommodated as a random network of clusters, chains and 2D patches of \cuzero. In what follows we discuss how the \cuzero network becomes reservoirs of resonating and strong singlets, via broad band Mott localization.

\section*{Emergence of singlet reservoirs in \cuzero subsystem}

Monovalent metals are special - they are half filled, very broad tight binding bands made of s like Wannier orbitals. Copper in bulk is a broad band 4s metal, well described by a half filled band of repulsive Hubbard model. However, 3 dimensionality makes Hubbard repulsive U irrelevant at low energy scales. A robust Fermi liquid state (which resits superconductivity, even at the lowest temperatures) wins and we get a jellium metal. Any electron correlation effects go off the energy shell \cite{GBAgAu1andIJMP,GBAgAu2019}

What is fascinating is that in low, namely zero and one dimensions low energy physics changes qualitatively and gives way to broad band Mott localization.\\

\textbf{Nano Clusters of \cuzero as Singlet Reservoirs}\\

Let us consider zero dimensions or small clusters of monovalent atoms for emergent strong pairing correlations. Monovalent metal atom clusters have been extensively studied from the point of view of electron correlation effects and magnetism, using Hubbard model for example \cite{TorriniZanazzi,Muhlschlegel,Pastor,NobleClusters,deHeerClusterRMP,PuruJena,HudaRay}. Nearest neighbor hoping t $\approx$ 3 eV and onsite repulsion U $\approx$ 7. For example Cu$_2$ exists as a stable molecule in free space as a tightly bound spin singlet with a binding energy of $\approx$ 2 eV. 

In small Cu clusters, large antiferromagnetic exchange J $\sim$ 2 eV gives rise to strong singlet bonds and resonances. In comparison to cuprates, where J $\approx$ 0.15 eV, we have an order of magnitude larger singlet pairing energy scale. Consequently local pairing temperature scale we get, on doping are several hundreds of Kelvin.

Effective Hamiltonian for a monovalent atom cluster is the Heisenberg antiferromagnetic Hamiltonian:
\be
H_c = \sum_{ij} J_{ij} ({\bf S}_i\cdot{\bf S}_j - 1/4) \equiv -\sum_{ij} J_{ij} 
b^\dagger_{ij}b^{}_{ij} 
\ee

Spin operator ${\bf S}_i \equiv d^{\dagger}_{i\alpha} {\hat\sigma}_{\alpha,\beta}d^{}_{j\beta}$.
Spin-spin interaction term has been rewritten using bond singlet operator
$b^\dagger_{ij} \equiv (1/\sqrt{2} (d^{\dagger}_{i\uparrow}d^{\dagger}_{j\downarrow} - d^{\dagger}_{i\downarrow}d^{\dagger}_{j\uparrow})$, (d's are electron operators) to emphasize the fact that strongly coupled bond singlets are natural basic units (rather than spin moments) in clusters. It is the resonance of these bond singlets (metal cluster aromaticity) which is at the heart of our mechanism for superconductivity. More importantly resonance energies of bond singlets are comparable to cohesion energy of monovalent clusters \cite{Pastor}.

Consider a simple tetrahedral cluster, an elementary closely packed unit in fcc lattice. Two singlets resonate and stabilize a singlet ground state. Larger clusters have been studied from the point of view of equilibrium shapes of clusters. For us what is important is the singlet content and their resonance. The tetrahedral RVB system has a doubly degenerate singlet ground state. The symmetric tetrahedra distorts via valence bond localization - this is an analogue of spin-Peierls distortion in the finite cluster.

We have suggested presence of a distribution of clusters in Cu Pb Apatite. As we go to larger clusters there are more complex resonances, depending on size and shape of a cluster. Singlets and their resonances continue to be present till a critical size. Beyond the critical size singlets and their resonances go off the energy shell \cite{GBAgAu1andIJMP,GBAgAu2019} and the bulk of the cluster behaves as a jellium metal (filled shells) at low energies. Surfaces of larger clusters are likely to have interesting singlet correlations and resonances.

It is clear that small clusters of various kinds are important as singlet reservoirs, as they provide a large singlet pairing energy scale J $\approx$ 2 eV.\\

\textbf{\cuzero Chains as Singlet Reservoirs}\\

In an ideal Cu chain in equilibrium, without dimerization, half filled band of 4s electrons get Mott localized \cite{LiebWu,AtomChainGoddard,MetalChainBook}, as a consequence of low dimensionality, even though the band is broad ($\sim 8$ eV wide) band. This spin-half Mott insulating chain becomes a reservoir of resonating singlets and large exchange energy scale J $\sim$ 2 eV. If we ignore lattice dimerazation instabilities, 1D Hubbard model predicts a Mott insulator with a Mott Hubbard gap of $\sim$ 1 eV (Lieb and Wu \cite{LiebWu}) and superexchange J $\sim$ 2 eV for Cu. Effective Hamiltonian of neutral Cu chain is given by an antiferromagnetic Heisenberg Hamiltonian similar to equation 1. 

When we move away from half filling we get, Luttinger liquid with spin-charge decoupling, enhanced local pairing and a very large fluctuationg pairing gap in ideal doped Cu chains. An organization of Cu chains into weakly interacting chain and doping easily brings in ambient Tc superconductivity as discussed below. 

For illustration, consider a 2D array of weakly coupled (for a range of interchain hopping t$_\perp$) Cu chains. To avoid competing antiferromagnetic 2 sublattice order, we stagger the chains, so that when t$_\perp$ = t we get isotropic triangular lattice. Using known studies  we find doped Mott insulator physics and ambient Tc superconductivity, provided competing instabilities such as dimerization and valence bond order are taken care of. 

In the above system pair tunneling plays an important role in stabilizing 2D or 3D superconducting order. We have used Wheatley-Hsu-Anderson's mechanism \cite{WHA} of interlayer pair tunneling and similar interchain pair tunneling theory of Zachar, Kivelson and Emery \cite{ZKEmery} to estimate superconducting Tc's.

Scale of Tc is given by k$_{\rm B}$Tc $\approx$ t$^2_\perp$/$\Delta_s$. Here $\Delta_s$ is a local spin gap scale. It can be approximated by superexchange J, in the present context. It is easily seen that for reasonable choice of t$_\perp$, in our model system of Ag and Au we get 2D superconducting Tc at ambient temperature scales. Thus we conclude that nanoscale Cu segments are also important singlet reservoirs, with a high pairing temperature scale.\\

\textbf{2D Planar \cuzero net as Singlet Reservoirs}\\

Two dimensions is equally interesting. We considered hypothetical 2D triangular lattices of Cu. Firstly, U is not strong enough to result in Mott localization. However, a form of (self) doped Mott insulator physics (dynamical Mott localization)  survives at and close to half filling. Using results from existing body of literature  \cite{TcThomale,TriangularTc1,TriangularTcTremblay,TriangularTcOhta,SumitMazumdar,Tanmoy,TcJarrel} we have estimated the Tc. After doping, results for Tc are not very different for triangular lattice and square lattice, even though symmetry of order parameters are different (d-wave for square lattice\cite{KotliarDwave}, and d + id for triangular lattice and honeycomb lattices \citep{GBCob,AnnaDoniach}).

Interestingly, in these model studies k$_B$Tc is a finite fraction of hoping parameter t, even for intermediate repulsions. In view of this non-BCS dependence we obtain ambient temperature (and higher) pairing temperature scales for Cu subsystem at and close to half filling.  

What is important for LK99 is that the nanoscale 2D patches will support strong singlet pairing,
which could eventually contribute to observed 3D hot Tc granular superconductivity.

\section*{Randomly Coupled Charged AMI's and Emergence of Granular Superconductivity}
As we mentioned earlier, weakly coupled reservoir of neutral singlets (clusters, chand and 2D patches) could transfer a fraction of 4s valence electrons to the insulating host. Transferred electrons will in general get trapped and become immobile, via disorder and correlation effects. In the charged AMI's resonating neutral singlets become charged Cooper pairs and get delocalized within the cluster, via physics of doped Mott insulator. Our clusters and chains, become Cooper pair boxes and are ready for Josephson coupling with neighbors. 

We have a model of randomly Josephson coupled Cooper pair boxes. In view of the strong local pairing scale, our estimate of superconducting Tc easily reach ambient temperature superconducting scales. We hope to discuss details in a later article.

\section*{Competing Orders}
Competing orders are ubiquitous in strongly correlated electron systems such as cuprates. These orders are typically charge and spin stripe orders accompanied by strong lattice distortions. Competing orders are encouraged when one has certain structural chemistry. For example, in LSCO, a monolayer cuprate, octahedral rotation strongly encourages stripe order and superconducting Tc comes down close to 30 K. Whereas in YBCO, a bilayer cuprate, the bilayer nature prevents easy octahedral rotation and Tc is more than 90 K. Trilayer cuprates, the middle layers are protected from competing orders and Tc's exceed 120 K.

In clusters and chains there is a strong tendency for dimerization type of instability. In chain it is the spin-Peierls (valence bond) order accommanied by lattice dimerization

\textit{It was our prediction \cite{GBAgAu1andIJMP} that monovalent single layer atomic net is a potential hot superconductor}. We also emphasize that in real 2D monoatomic monovalent metal system \cite{2DMetalAtlas}, when synthesized, competing orders such as valence bond order and lattice dimerization will compete and supress superconductivity. Free hanging 2D layers are good bet, as they are not influenced by 2D layer grown on substrates.

A deeper understanding of structural chemistry will enable structural design with reduced competing order.

\section*{Discussion and Conclusion} 
Excited by the report of observation of hot superconductivity in the Cu-Pb Apatite system, we have presented a couple of scenarios, keeping solid state and quantum chemical constraints in mind and suggested possibility of hot superconductivity via electronic mechanism of doped Spin\half Mott insulator, utilizing broad band Mott localization.

Our theory makes use of doped  spin-\half doped Mott insulator route to superconductivity, with the emphasis on the fact that  monovalent metal atoms under suitable conditions (such as reduced dimensions) readily undergo Mott localization. Over years we have used this AMI idea to develop tangible models to understand superconductivity in complex solid state systems such as hydrides, K$_{3}$p-Terphenyl, B-doped diamond and Ag-Au nanostructures, doped graphene and silicene.

In the present article we have proposed a compelling scenario, without full justification from quantum chemical calculations etc. In the same spirit, we briefly discuss possibility of a related, but somewhat different scenario. Here Cu atom get accommodated as a neutral impurity and creates an s-like donor impurity state in the band gap of \pbap. As impurity concentration is increased impurity state wave functin will overlap and we could have a Anderson-Mott insulator to superconductor transition, as outlined in our theory for B doped diamond \cite{GBDopedDiamond}. Here also, depending on the binding energy of the donor state we could in principle get Tc reaching the scales claimed.

As a key idea and notion, superconductivity in doped Mott insulators began while attempting to understand high Tc cuprates by Anderson in 1987. It is becoming increasingly clear that this idea has a wide applicability. We once again emphasize, excited by our theory, experimental exploration of the rich field of low dimensional monovalent metals, monovalent metal doped minerals, insulators and in metallurgy, where variety of nanostructures are norm of the day.

We also wish to reiterate that while \cu2plus doping is an attractive idea, correponding isolated bare band widths are so low that Cooper pairs will get localized strongly by lattice distortion and disorder. For the scale of reported Tc, we need broad band Mott localization emphasized in the present article, rather than narrow band Mott localization. It is important to recall that even in high Tc cuprates an effective single band (which is not isolated) width is of the order of 2 eV, which is capable of giving superconductivity. at temperatures close to 100 K.

Our broad band Mott localization and AMI is likely to play a key role in other systems perhaps. We conclude our article by reproducing a paragraph from  our 2019 arXiv article on broad band Mott localization and emergent AMI (reference \cite{GBAgAu2019}, page 6, 2nd column, 3rd paragraph in discussion session): 

\textit{`Does ambient temperature granular superconductivity exist already, in some form or other, in the rich world of minerology and metallurgy - natural ones and man made ? An exploration into minerology and metallurgy is likely to offer surprises'}.
 
\section*{Acknowledgement}
I thank Veer Awana for sharing his very recent results on synthesis and measurements of LK-99 samples. Discussions with V N Muthukumar, after several years, on issues of hot superconductivity and renewal of our collaboration is acknowledge. Brief but encouraging discussions with Venky Venkatesan and Rajaram Nityananda are acknowledged. Valuable help from Bala Yogendra for arXiv uploading is acknowledged. I thank the Institute of Mathematical Sciences, Chennai, IITMadras and Perimeter Institute of Theoretical Physics, Waterloo, Canada for continuing hospitality. Perimeter Institute for Theoretical Physics, Waterloo, Canada is supported by the Government of Canada through Industry Canada and by the Province of Ontariothrough the Ministry of Research and Innovation.

P.W. Anderson showed us a fertile direction to achieve hot supercondtivity. A phenomenologist par excellence, who listened to nature always, Phil was skeptical about existence of hot superconductors, before cuprates. In his article \cite{PWAPastFuture} \textit{Superconductivity in the Past and Future} (1969) he states, `Room-temperature magnets or superconducting skis ? Probably not. For one thing, so many diverse types of materials exist in nature or have been made in the laboratory that large, obvious effects would probably already have been seen. But room-temperature interferometers ? Who knows ?' 

My obsession with possible hot superconductivity, via broad band Mott localization, ever since I got entangled with RVB theory from January 1987, did bring constant appreciation from Phil. I dedicate this article to the memory of P.W. Anderson, my mentor, whose 100th birth anniversary is on the horizon.


\begin{thebibliography}{0}%
\makeatletter
\providecommand \@ifxundefined [1]{%
 \@ifx{#1\undefined}
}%
\providecommand \@ifnum [1]{%
 \ifnum #1\expandafter \@firstoftwo
 \else \expandafter \@secondoftwo
 \fi
}%
\providecommand \@ifx [1]{%
 \ifx #1\expandafter \@firstoftwo
 \else \expandafter \@secondoftwo
 \fi
}%
\providecommand \natexlab [1]{#1}%
\providecommand \enquote  [1]{``#1''}%
\providecommand \bibnamefont  [1]{#1}%
\providecommand \bibfnamefont [1]{#1}%
\providecommand \citenamefont [1]{#1}%
\providecommand \href@noop [0]{\@secondoftwo}%
\providecommand \href [0]{\begingroup \@sanitize@url \@href}%
\providecommand \@href[1]{\@@startlink{#1}\@@href}%
\providecommand \@@href[1]{\endgroup#1\@@endlink}%
\providecommand \@sanitize@url [0]{\catcode `\\12\catcode `\$12\catcode
  `\&12\catcode `\#12\catcode `\^12\catcode `\_12\catcode `\%12\relax}%
\providecommand \@@startlink[1]{}%
\providecommand \@@endlink[0]{}%
\providecommand \url  [0]{\begingroup\@sanitize@url \@url }%
\providecommand \@url [1]{\endgroup\@href {#1}{\urlprefix }}%
\providecommand \urlprefix  [0]{URL }%
\providecommand \Eprint [0]{\href }%
\providecommand \doibase [0]{https://doi.org/}%
\providecommand \selectlanguage [0]{\@gobble}%
\providecommand \bibinfo  [0]{\@secondoftwo}%
\providecommand \bibfield  [0]{\@secondoftwo}%
\providecommand \translation [1]{[#1]}%
\providecommand \BibitemOpen [0]{}%
\providecommand \bibitemStop [0]{}%
\providecommand \bibitemNoStop [0]{.\EOS\space}%
\providecommand \EOS [0]{\spacefactor3000\relax}%
\providecommand \BibitemShut  [1]{\csname bibitem#1\endcsname}%
\let\auto@bib@innerbib\@empty
\end{thebibliography}%


\begin{thebibliography}{99}
	
\bibitem{BednorzMuller1986} G. Bednorz and K.A. Muller, Z. Phys. \textbf{B 64} 189 (1986)

\bibitem{PWA1987} P.W. Anderson, Science \textbf{235} 1196 (1987)

\bibitem{BZAandOthers} G. Baskaran, Z. Zou and P.W. Anderson, Solid St. Com. \textbf{63} 973 (1987); P.W. Anderson, G. Baskaran, Z. Zou and T. Hsu,
Phys.Rev.Lett., \textbf{58} 2790 (1987); G. Baskaran and P.W. Andersdon, Phys. Rev., \textbf{B 37} 580 (1988)

\bibitem{KRS} S.A. Kivelson, D.S. Rokhsar, and J. P. Sethna
Phys. Rev., \textbf{B 35} 8865(R) (1987) 

\bibitem{KotliarDwave} G. Kotliar, Phys. Rev. \textbf{B 37} 3664 (1988)

\bibitem{PlainVanila} P.W. Anderson et al., J. Physics: Condensed Matter, \textbf{16} R755 (2004)

\bibitem{GBIran} G. Baskaran, Iranian J. Phys. Res. \textbf{6} 163 (2006); arXiv:cond-mat:0611548

\bibitem{WenBook} X.G. Wen, Quantum Field Theory of Many-Particle Systems
(Oxford University Press 2010)

\bibitem{TaiKaiSpinLiquid} Y. Zhou, K. Kanoda T.K. Ng, Rev. Mod. Phys., \textbf{89} 025003 (2017)

\bibitem{GB5FoldWay} G. Baskaran, Pramana, \textbf{73} 61 (2009);\\
https://www.ias.ac.in/article/fulltext/pram/073/01/0061-0112

\bibitem{DiasLuNH} N. Dasenbrock-Gammon et al., Nature, \textbf{615} 244 (2023)

\bibitem{LaHx400} A.D. Grockowiak et al., Frontiers in Electronic Materials, \textbf{2} 2.837651 (2022)

\bibitem{ThapaPandey} D.K.Thapa  arXiv:1807.08572; 
S.K. Saha et al 2022 Supercond. Sci. Technol. \textbf{35} 084001 (2022)

\bibitem{LK99} Sukbae Lee, Ji-Hoon Kim and Young-Wan Kwon, arXiv:2307.12008; Sukbae Lee et al., arXiv:2307.12037

\bibitem{Kopelevich}Y. Kopelevich, R.R. da Silva and B.C. Camargo,
Physica \textbf{C 514} 237 (2015); Y. Kopelevich and P.D. Esquinazi,
J. Low Temp. Phys. \textbf{146} 629 (2007); P.D. Esquinazi et al., Quantum Studies: Mathematics and Foundations, \textbf{5} 41 (2018)

\bibitem{GBMgB2} G. Baskaran, Phys. Rev. \textit{B65} 212505 (2002)

\bibitem{AnnaDoniach} Annica M. Black-Schaffer and Sebastian Doniach,
Phys. Rev.\textbf{B75} 134512 (2007)

\bibitem{PathakShenoyGB} S. Pathak, V.B. Shenoy and G. Baskaran, Phys. Rev. \textbf{B 81} 085431 (2010)

\bibitem{GBSilicene} G. Baskaran, arXiv:1309.2242 (2013); Book Chapter, Many-body Approaches in Different Scales (Eds. G.G.N. Angilella and C. Amovilli; Springer Nature 2018)

\bibitem{GBDopedDiamond} G. Baskaran, arXiv:cond-mat:0404286 (2004); Sci. Technol. Adv. Mater. \textit{7} S49 (2006); Sci. Technol. Adv. Mater. \textbf{9} 044104 (2008)

\bibitem{GBpTerphenyl} G Baskaran, arXiv:1704.08153 (2017) 

\bibitem{GBHydride} G. Baskaran, arXiv:1507.03921 (2015)

\bibitem{GBAgAu2019} G. Baskaran, arXiv:1906.02143

\bibitem{GBAgAu1andIJMP} G. Baskaran, Int. J. Mod. Phys. \textbf{B36} 2250184 (2022); arXiv:1808.02005 (2018)

\bibitem{WHA} J. Wheatley, T. Hsu and P.W. Anderson, Nature, \textbf{333} 121 (1988); V.N. Muthukumar and G. Baskaran, Mod. Phys. Lett. \textbf{B8} 699 (1994); G. Baskaran, Phil. Mag., \textbf{B76} 119 (1997)

\bibitem{ZKEmery} Z. Zachar, S.A. Kivelson and V. J. Emery, Journal of Superconductivity, \textbf{10} 373 (1997)

\bibitem{CuDiaMSRao} S. Ramachandra Rao, Nature, \textbf{136} 436 (1935) 

\bibitem{ImryDiamagnetism} J. Imry, Phys. Rev., \textbf{B 91} 104503 (2015)

\bibitem{AuDiaNanorod} A Hernando et al., New Journal of Physics \textbf{16} 073043 (2014)

\bibitem{AuDiaNanorodRhee} P.G. van Rhee et al., Phys. Rev. Lett., \textbf{111} 127202 (2013)

\bibitem{AgCuCompositeDia} G. Frommeye and K.F. Schneider, Phys. Stat. Sol:\textbf{(a)41} 653 (1977)

\bibitem{AgRingDia} R. Deblock et al., Phys. Rev. Lett., \textbf{89} 206803 (2002)

\bibitem{AgAuFM} W. Luo et al., Nano Letters, \textbf{7} 3134 (2007); V Tuboltsev et al., ACS Nano, \textbf{7-8} 6691 (2013)

\bibitem{LK99Abinitio} Junwen Lai et al., arXiv:2307.16040; S.M. Griffin, arXiv:2307.16892; Liang Si and Karsten Held, arXiv:2308.00676; Rafal Kurleto et al., arXiv:2308.00698

\bibitem{LK99Awana} Kapil Kumar, N.K. Karn and V.P.S. Awana, arXiv:2307.16402
\bibitem{LK99Chinese} Li Lu et al., arXiv:2307.16802

\bibitem{ApatiteIonicCond} Xiaoyan Yang, Alberto J. Fernández-Carrión and Xiaojun Kuang, Chem. Rev., (24th July 2023), online: https://doi.org/10.1021/acs.chemrev.2c00913

\bibitem{TorriniZanazzi} M Torrini and E Zanazzi, J. Phys. C: Solid State Phys.\textbf{9}  63 (1976)

\bibitem{Muhlschlegel} G. M. Pastor, R. Hirsch and B. Muhlschlegel, Phys. Rev. \textbf{B 53} 10382 (1996)

\bibitem{Pastor} G.M. Pastor in Atomic clusters and nanoparticles (Ed.C. Guet et al.) Les Houces Session LXXIII (Springer 2000)

\bibitem{NobleClusters} Eva M. Fernández et al., Phys. Rev. \textbf{B 70} 165403 (2004)

\bibitem{deHeerClusterRMP} W.A. de Heer, Rev. Mod. Phys., \textbf{65} 611 (1993)

\bibitem{PuruJena} B. K. Rao and P. Jena, Phys. Rev.,\textbf{ B32} 2058 (1985)

\bibitem{HudaRay} M.N. Huda and A.K. Ray, Eur. Phys. J., \textbf{D 22} 217 (2003)

\bibitem{LiebWu} E. Lieb and F.Y. Wu, Phys. Rev. Lett., \textbf{20} 1445 (1968)

\bibitem{AtomChainGoddard}M. H. McAdon and W.A. Goddard III, The J. of Chem. Phys., \textbf{88} 277 (1988)

\bibitem{MetalChainBook} Metal Chain/Chains of Metal, Ed. M. Springborg and 
Yi Dong (Elsevier 2007)

\bibitem{TcThomale} M. Laubach et al., Phys. Rev., \textbf{B 91} 245125 (2015)

\bibitem{TriangularTc1} J. Wang et al., Int. J. Mod. Phys. \textbf{C28} 1750127 (2017)

\bibitem{TriangularTcTremblay} C-D. Hebert, Semon and A.-M.S. Tremblay, Phys. Rev.,\textbf{B 92} 195112 (2015); S. Acheche et al., Phys. Rev., \textbf{B 94} 245133 (2016)

\bibitem{TriangularTcOhta} K. Misumi, T. Kaneko, and Y. Ohta, Phys. Rev., \textbf{B 95} 075124 (2017)

\bibitem{SumitMazumdar} W.W.De Silva et al., Phys. Rev., \textbf{B 93} 
205111 (2016)

\bibitem{Tanmoy} Tanmoy Das, R.S. Markiewicz and A. Bansil (2014) Advances in Physics, \textbf{63} 151 (2014) DOI: 10.1080/00018732.2014.940227

\bibitem{TcJarrel} K.-S. Chen et al., Phys. Rev. \textbf{B 88} 245110 (2013)

\bibitem{GBCob} G. Baskaran, Phys. Rev. Lett. \textbf{91} 097003 (2003)

\bibitem{2DMetalAtlas} Janne Nevalaita and Pekka Koskinen, Phys. Rev. \textbf{B97} 035411 (2018)
 
\bibitem{PWAPastFuture} P.W. Anderson, in Superconductivity, pp. 1343-1358 Part 2, edited by R. Parks
(Marcel Dekker, New York, 1969); this article is reproduced in P.W. Anderson's book `A Career in Theoretical Physics' (World Scientific, 2004)



\end{thebibliography}
\end{document}